\documentclass[aps,preprint]{revtex4}
\usepackage{amsmath,amssymb,amsthm,graphicx,latexsym}
\newcommand{\bd}{\begin{document}}
\newcommand{\ed}{\end{document}}
\newcommand{\bc}{\begin{center}}
\newcommand{\ec}{\end{center}}
\newcommand{\bfr}{\begin{flushright}}
\newcommand{\efr}{\end{flushright}}
\newcommand{\lt}{\left}
\newcommand{\rt}{\right}
\newcommand{\vs}{\vspace}
\newcommand{\hs}{\hspace}
\newcommand{\beq}{\begin{equation}}
\newcommand{\eeq}{\end{equation}}
\newcommand{\lb}{\linebreak}
\newcommand{\pb}{\pagebreak}
\newcommand{\mb}{\makebox}
\newcommand{\fb}{\framebox}
\newcommand{\mc}{\multicolumn}
\newcommand{\ben}{\begin{enumerate}}
\newcommand{\een}{\end{enumerate}}
\newcommand{\bit}{\begin{itemize}}
\newcommand{\eit}{\end{itemize}}
\newcommand{\ovl}{\overline}
\newcommand{\un}{\underline}
\newcommand{\lefq}{\lefteqn}
\newcommand{\ba}{\begin{array}}
\newcommand{\ea}{\end{array}}
\newcommand{\beqa}{\begin{eqnarray}}
\newcommand{\eeqa}{\end{eqnarray}}
\newcommand{\beqas}{\begin{eqnarray*}}
\newcommand{\eeqas}{\end{eqnarray*}}
\newcommand{\bfg}{\begin{figure}}
\newcommand{\efg}{\end{figure}}
\newcommand{\bds}{\begin{displaymath}}
\newcommand{\eds}{\end{displaymath}}
\newcommand{\btb}{\begin{tabbing}}
\newcommand{\etb}{\end{tabbing}}
\newcommand{\para}{\parallel}
\newcommand{\pad}{\partial}
\newcommand{\nn}{\nonumber}
\newcommand{\la}{\leftarrow}
\newcommand{\ra}{\rightarrow}
\newcommand{\lgla}{\longleftarrow}
\newcommand{\lgra}{\longrightarrow}
\newcommand{\La}{\Leftarrow}\newcommand{\Ra}{\Rightarrow}
\newcommand{\Lra}{\Leftrightarrow}
\newcommand{\Lgla}{\Longleftarrow}
\newcommand{\Lgra}{\Longrightarrow}
\newcommand{\bm}{\boldmath}
\newcommand{\lan}{\langle}
\newcommand{\ran}{\rangle}
\renewcommand{\a}{\alpha}
\renewcommand{\b}{\beta}
\newcommand{\g}{\gamma}
\newcommand{\G}{\Gamma}
\renewcommand{\d}{\delta}
\newcommand{\eps}{\epsilon}
\newcommand{\s}{\sigma}
\newcommand{\D}{\Delta}
\newcommand{\vare}{\varepsilon}
\newcommand{\pr}{\prime}
\newcommand{\ro}{\rho}
\newcommand{\nab}{\nabla}
\newcommand{\m}{\mu}
\newcommand{\n}{\nu}
\newcommand{\Sg}{\Sigma}
\newcommand{\p}{\pi}
\newcommand{\R}{I\!\!R}
\newcommand{\om}{\omega}
\newcommand{\Om}{\Omega}
\newcommand{\ze}{\zeta}
\newcommand{\vart}{\vartheta}
\newcommand{\lam}{\lambda}
\newcommand{\tri}{\triangle}
\newcommand{\f}{\frac}
\newcommand{\iny}{\infty}
\newcommand{\pro}{\propto}
\newcommand{\np}{\newpage}
\newcommand{\ds}{\displaystyle}
\begin{document}
\title{\large  An analysis of the zero energy states in graphene}

\author{\textsc{P.~Ghosh}}
\affiliation{Department of Mathematics, Binapani Balika Vidyalaya (HS), North 24 Parganas 743424, India.\\}
\author{\textsc{P.~Roy}}
\email[Email: ]{pinaki@isical.ac.in}
\affiliation{Physics and Applied Mathematics Unit, Indian Statistical Institute, Kolkata 700 108, India}

\vspace{2cm}

\begin{abstract}
Using the concept of complex non $\cal{PT}$ symmetric potential we study creation of zero energy states in graphene by a scalar potential. The admissible range of the potential parameter values for which such states exist has been examined. The situation with respect to the holes has also been investigated.
\end{abstract}
\maketitle
\newpage
Bound states play an important role in the context of controlling the motion of electrons in graphene. Bound states in graphene can be created using either magnetic fields \cite{m1,m2,m3} or scalar potentials \cite{e1,e2,hart}. In particular, zero energy states in graphene created by using magnetic fields \cite{bm1,bm2,bm3,bm4,bm5} as well as scalar potentials \cite{rob,dow,hr} have been studied. Here we shall propose a new potential to confine the electrons (and the holes) and examine in detail the corresponding zero energy states. It will be seen that after diagonalization the components obey Schr\"odinger like equations with complex potentials which have real spectrum but are not $\cal{PT}$ symmetric \cite{bender,ahmed,z,cmb}. Also the examples considered here would illustrate that complex potentials possessing real spectrum and without $\cal{PT}$ symmetry \cite{real1,real2,real3} are not merely of theoretical interest but they can be related to Hermitian systems. Furthermore it will be seen that one may reproduce the results of a previous study \cite{rob,hr} by suitably choosing the potential parameters.

The motion of electrons in graphene in the presence of a potential is governed by the equation
\beq\label{1}
[v_F(\sigma_xp_x+\sigma_yp_y)]\psi+U(x,y)\psi=E\psi
\eeq
where $v_F=10^6~m/s$ is the Fermi velocity, $(\sigma_{x},\s_y)$ are the Pauli spin matrices and $U(x,y)$ is a potential. In what follows we shall consider a potential depending on the $x$ coordinate only and so the wavefunction can be taken as
\beq\label{wf}
\psi(x,y)=e^{ik_yy}\left(\ba{c}\psi_A(x) \\ \psi_B(x)\ea\right).
\eeq
Now substituting (\ref{wf}) in Eq.(\ref{1}) we obtain
\beq\label{intert1}
\ds (V(x)-\eps)\psi_A-i\left(\f{d}{dx}+k_y\right)\psi_B=0
\eeq
\beq\label{intert2}
\ds (V(x)-\eps)\psi_B-i\left(\f{d}{dx}-k_y\right)\psi_A=0
\eeq
where $V(x)=U(x)/\hbar v_F$ and $\eps=E/\hbar v_F$. 

It may be noted that Eqs.~(\ref{intert1}) and (\ref{intert2}) are invariant under the following transformations:
\beq\label{ky}
k_y\to -k_y,~~ \psi_A\leftrightarrow \psi_B
\eeq
From Eq.(\ref{ky}) it follows that if the spinor $\psi=e^{ik_yy}(\psi_A,\psi_B)^t$ is a solution for $k_y$, then $\psi=e^{-ik_yy}(\psi_B,\psi_A)^t$ is also a solution for $-k_y$.

It is now necessary to disentangle the components $\psi_{1,2}$. To this end we define 
\beq
\psi_{1,2}(x,y)=(\psi_A(x)\pm\psi_B(x))
\eeq
and obtain from Eqs.(\ref{intert1}) and (\ref{intert2})
\beq\label{intert3}
\ds\left(V(x)-\eps-i\f{d}{dx}\right)\psi_1+ik_y\psi_2=0
\eeq
\beq\label{intert4}
\ds\left(V(x)-\eps+i\f{d}{dx}\right)\psi_2-ik_y\psi_1=0
\eeq
From Eqs.(\ref{intert3}) and (\ref{intert4}) it can be easily shown that the components $\psi_{1,2}$ satisfy the following equations
\beq\label{5}
\left[-\f{d^2}{dx^2}-(V(x)-\eps)^2-i \f{dV}{dx}+k_y^2\right]\psi_1=0
\eeq
\beq\label{6}
\left[-\f{d^2}{dx^2}-(V(x)-\eps)^2+i \f{dV}{dx}+k_y^2\right]\psi_2=0
\eeq

Eqs.(\ref{5}) and (\ref{6}) are Schr\"odinger like equations and for zero energy ($\eps=0$) the potentials in these equations become 
\beq\label{pot3}
V_1(x)=-V^2(x)-i\f{dV}{dx}+k_y^2,
\eeq
\beq\label{pot4}
V_2(x)=-V^2(x)+i\f{dV}{dx}+k_y^2.
\eeq
It is interesting to note that the above potentials are complex for real $V(x)$ and in case $V(x)$ is an even function the potentials are $\cal{PT}$ symmetric. 

We now consider a potential which has not been considered before and is given by
\beq\label{pot}
V(x)=-\lam sechx + \mu tanhx,~~~~-\infty<x<\infty
\eeq
where $\lam$ and $\mu$ are real constants. It may be noted that (\ref{pot}) is a potential well for electrons when $\lam>0$  while for $\lam<0$ it is a potential well for the holes. Now using (\ref{pot}) in Eqs.(\ref{pot3}) and (\ref{pot4}) the effective potentials $V_{1,2}(x)$ are found to be
\beq\label{v12}
V_{1,2}(x)=(\mu^2\mp i\mu-\lam^2)sech^2x+\lam(2\mu\mp i)sechx~tanhx+k_y^2-\mu^2
\eeq
The above potentials are of the form of complex Scarf II potential
\beq\label{compare}
U_S(x)=-(B^2+A^2+A)~sech^2x + iB(2A+1)~sechx~tanhx
\eeq
Depending on the nature of the constants $A,B$ the potential $U_S(x)$ may or may not be $\cal{PT}$ symmetric. It may be recalled that $\cal{PT}$ potentials admit real eigenvalues. However in a recent interesting paper it has been shown \cite{nathan} that even when the potential $U_S(x)$ is {\it not $\cal{PT}$ symmetric} it may still possess a real spectrum. Before we discuss the effective potentials $V_{1,2}$ we briefly mention some results concerning complex Scarf II potential \cite{ahmed,khare,nathan}. The Schr\"odinger equation for $U_S(x)$
\beq
-\f{d^2\phi}{dx^2}+U_S(x)\phi=E\phi,
\eeq
is exactly solvable and the solution relevant for our purpose is given by 
\beq\label{en1}
E_n=-(A-n)^2,~~n=0,1,2,....<A
\eeq
\beq\label{phi1}
\phi_{n}(x)=(sechx)^A~exp[-iBtan^{-1}(sinhx)]~P_{n}^{(-A+B-\f{1}{2},-A-B-\f{1}{2})}(isinhx)
\eeq
A feature of the complex Scarf II potential (\ref{compare}) is that it is invariant under the substitution $A+\f{1}{2}\leftrightarrow B$.  Consequently there exists a second set of solutions given by 
\beq
E_n=-\left(n-B+\f{1}{2}\right)^2,~~n=0,1,2,....<B-\f{1}{2}
\eeq
\beq\label{phi2}
\phi_{n}(x)=(sechx)^{(B-\f{1}{2})}~exp[-i(A+\f{1}{2})tan^{-1}(sinhx)]~P_{n}^{(A-B+\f{1}{2},-A-B-\f{1}{2})}(isinhx)
\eeq
We now turn to the effective potentials $V_{1,2}$. Clearly if the solutions of one of them are known then the solution for the other may be obtained by using the intertwining relations (\ref{intert3}) or (\ref{intert4}). To be specific, let us consider $V_1(x)$. By comparing $V_1(x)$ with $U_S(x)$ we find the following possibilities :
\beq\label{AB}
\ba{l}
(a)~~A=\lam-\f{1}{2},~~~~B=-\f{1}{2}-i\mu,~~~~(b)~~A=-\lam-\f{1}{2},~~~~B=\f{1}{2}+i\mu\\
(c)~~A=i\mu,~~~~B=-\lam,~~~~(d)~~A=-1-i\mu,~~B=\lam
\ea
\eeq
\newpage

{\bf Case 1. $\lam>0$.}

From (\ref{phi1}) it is seen that for normalizable solutions $A>0$. This requires $\lam>\f{1}{2}$ and the correct choice is given by $(a)$ above. The solutions can now be obtained from (\ref{en1}) and (\ref{phi1}) and are given by
\beq\label{wf1}
\ba{l}
 \eps=0,~~~~k_y=\pm\sqrt{\mu^2+(\lam-\f{1}{2}-n)^2},~~n=0,1,2,....<\lam-\f{1}{2}\\
\psi_{1,n}(x)=(sechx)^{\lam-\f{1}{2}}~exp[(-\mu+\f{i}{2})tan^{-1}(sinhx)]~P_{n}^{(-\lam-i\mu-1/2,-\lam+i\mu+1/2)}(isinhx)
\ea
\eeq
Let us now examine the eigenfunctions of the Dirac problem. As mentioned earlier one may obtain $\psi_2(x)$ by directly solving the Schr\"odinger equation for $V_2(x)$. However one must also ensure that this solution satisfies the intertwining relations (\ref{intert3}) and (\ref{intert4}). Thus using (\ref{wf1}) in (\ref{intert3}) we obtain after some calculations
\beq
\psi_{2,n}=i\f{(n-\lam+i\mu+\f{1}{2})}{(\lam-n-\f{1}{2})}~(sechx)^{(\lam-\f{1}{2})}exp[-(\mu+\f{i}{2})tan^{-1}(sinhx)]~P_n^{(-\lam-i\mu+\f{1}{2},-\lam+i\mu-\f{1}{2})}(isinhx)
\eeq
So, finally the zero energy solutions of the original Dirac problem are given by
\beq\label{final>0}
\ba{l}
\eps=0,~~~~k_y=\pm\sqrt{\mu^2+(\lam-\f{1}{2}-n)^2},~~n=0,1,2,....<\lam-\f{1}{2}\\
\psi_{A(B),n}=\f{1}{2}(sechx)^{(\lam-\f{1}{2})}exp[-(\mu+\f{i}{2})tan^{-1}(sinhx)]\times\\
~\left[exp[i~tan^{-1}(sinhx)]P_n^{(-\lam-i\mu-\f{1}{2},-\lam+i\mu+\f{1}{2})}(isinhx)\pm i\f{(n-\lam+i\mu+\f{1}{2})}{(\lam-n-\f{1}{2})}P_n^{(-\lam-i\mu+\f{1}{2},-\lam+i\mu-\f{1}{2})}(isinhx)\right]
\ea
\eeq 
 Two points are to be noted here. First, the second set of solutions do not exist as the transformation $A+\f{1}{2}\leftrightarrow B$ corresponds to the choice $(d)$ for which the wavefunctions are non normalizable. Secondly, the solutions in (\ref{final>0}) are valid for $\lam>\f{1}{2}$ and no solutions exist in the range $0<\lam\leq \f{1}{2}$. 

{\bf Case 2. $\lam<0$.}

In this case we are treating bound state solutions for the holes. The appropriate choice of parameters is given by $(b)$ and this requires $\lam<-\f{1}{2}$. These solutions can be obtained from (\ref{phi1}) and are given by
\beq\label{a-ve}
\ba{l}
E_n=-(n+\lam+\f{1}{2})^2,~~~~n=0,1,2,....<-\lam-\f{1}{2} \\
\phi_n(x)=(sechx)^{-(\lam+\f{1}{2})}~exp[(\mu-\f{i}{2})tan^{-1}(sinhx)]~P_n^{(\lam+i\mu+\f{1}{2},\lam-i\mu-\f{1}{2})}(isinhx)
\ea
\eeq

Thus the solutions of the Dirac problem for $\lam<-\f{1}{2}$ are given by
\beq\label{final1<0}
\ba{l}
\eps=0,~~~~k_y=\pm\sqrt{\mu^2+(\lam+n+\f{1}{2})^2},~~~~n=0,1,2,....<-\lam-\f{1}{2} \\
\psi_{A(B),n}=\f{1}{2}(sechx)^{-(\lam+\f{1}{2})}exp[(\mu+\f{i}{2})tan^{-1}(sinhx)]\times\\
~\left[exp[-i~tan^{-1}(sinhx)]P_n^{(\lam+i\mu+\f{1}{2},\lam-i\mu-\f{1}{2})}(isinhx)\pm \f{(n+\lam+i\mu+\f{1}{2})}{i(\lam+n+\f{1}{2})}P_n^{(\lam+i\mu-\f{1}{2},\lam-i\mu+\f{1}{2})}(isinhx)\right]
\ea
\eeq 
As in the previous case here also bound state solutions do not exist in the region $-\f{1}{2}\leq \lam<0$ and also the transformation $A+\f{1}{2}\leftrightarrow B$ do not yield any normalizable wavefunction.

We shall now consider another potential which has been considered before \cite{rob,hr} and is obtained by putting $\mu=0$ in (\ref{pot}) :
\beq\label{pot1}
{\tilde V}(x)=-\lam sechx,~~~~\lam>0
\eeq
For the purpose of comparison we have presented a plot of the potentials (\ref{pot}) and (\ref{pot1}) in Fig 1. From Fig 1 it is seen that both are single well potentials and can be made very similar/dissimilar by suitably choosing the parameters. Clearly we may obtain the solutions of the Dirac equation with this potential by putting $\mu=0$ in the previous ones. There are two points which need mention. First, the potential (\ref{pot1}) is an even function and consequently the effective potentials (obtained by putting $\mu=0$ in (\ref{v12})) are $\cal{PT}$ symmetric. Secondly, in this case also bound state solutions can not be found in the range $-\f{1}{2}\leq\lam\leq \f{1}{2}$.

In conclusion we have studied zero energy states of electrons as well as holes in graphene using a hyperbolic potential. Interestingly this potential leads to a pair of Schr\"odinger equation with {\it non $\cal{PT}$ symmetric} potentials with real eigenvalues. Also we can obtain the solutions for a potential studied before by putting $\mu=0$. Furthermore the potential studied here serves as an illustration of the use of complex nonrelativistic systems without $\cal{PT}$ symmetry. In this context it may be noted that since we started from a Hermitian relativistic system given by Eq.(\ref{1}) with a real potential $U(x)$ the eigenvalues must necessarily be real. However the method of solution leads to a pair of non $\cal{PT}$ symmetric effective potentials with real spectrum. Thus this approach may be considered as a method to construct general complex potentials with real spectrum. Finally we remark that the potential $\tilde{V}(x)$ has been shown to be quasi exactly solvable \cite{hart} and we believe it would be of interest to examine quasi exact solvability of $V(x)$ as well as to find other potentials admitting exact zero energy solutions.
      
\begin{figure}[h]
	\centering
		\includegraphics{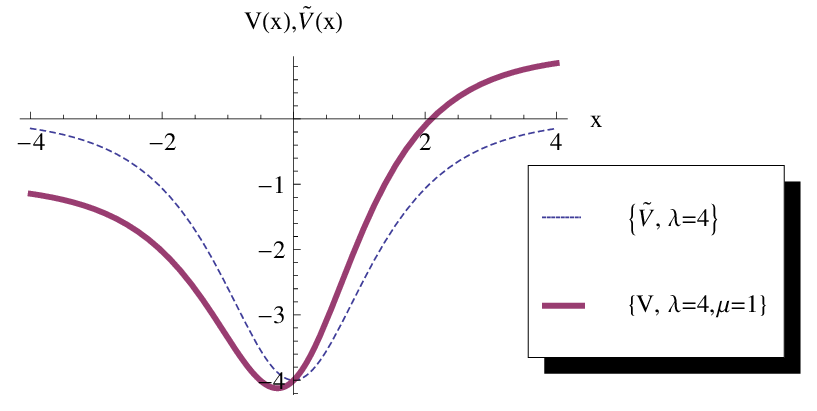}
		\caption{Graphs of the potentials $V(x)$ and $\tilde{V}(x)$ for $\lam=4, \mu=1$.}
\end{figure}

\np
\bc
{\bf Acknowledgement}
\ec
One of the authors (PR) wishes to thank Zafar Ahmed for many helpful discussions.

\ed